\documentclass{jaa}
\usepackage{natbib}
\usepackage{xcolor}
\usepackage{multicol}

\usepackage{hyperref}
\hypersetup{
    colorlinks=true,
    linkcolor=blue,
    filecolor=magenta,      
    urlcolor=cyan,
    pdftitle={Overleaf Example},
    pdfpagemode=FullScreen,
    citecolor=red,
    }

\usepackage{graphicx}
\usepackage{url}

\begin{document}\sloppy

\title{Progression of Digital-Receiver Architecture: from MWA to SKA1-Low, and beyond}
\author{
Girish B. S.\textsuperscript{1,*},   
Harshavardhan Reddy S.\textsuperscript{3},
Shiv Sethi\textsuperscript{2},
Srivani K. S.\textsuperscript{1},   
Abhishek R.\textsuperscript{1},
Ajithkumar B.\textsuperscript{3},
Sahana Bhattramakki\textsuperscript{1},
Kaushal Buch\textsuperscript{3},
Sandeep Chaudhuri\textsuperscript{3},
Yashwant Gupta\textsuperscript{3},
Kamini P. A.\textsuperscript{1},
Sanjay Kudale\textsuperscript{3},
Madhavi S.\textsuperscript{1},
Mekhala Muley\textsuperscript{3},
Prabu T. \textsuperscript{1},
Raghunathan A.\textsuperscript{1} and
Shelton G. J.\textsuperscript{3}
}

\affilOne{\textsuperscript{1}Electronics Engineering Group, Raman Research Institute, Bengaluru-560080, India.{}\\}
\affilTwo{\textsuperscript{2}Astronomy \& Astrophysics, Raman Research Institute, Bengaluru-560080, India.{}\\}
\affilThree{\textsuperscript{3}Giant Metrewave Radio Telescope, NCRA-TIFR, Pune-411007, India.{}}


\twocolumn[{

\maketitle

\corres{bsgiri@rri.res.in}

\msinfo{X March 2022}{X YYY ZZZZ}

\begin{abstract}
Backed by advances in digital electronics, signal processing, computation, and storage technologies, aperture arrays, which had strongly influenced the design of telescopes in the early years of radio astronomy, have made a comeback. Amid all these developments, an international effort to design and build the world’s largest radio telescope, the Square Kilometre Array (SKA), is ongoing. With its vast collecting area of 1 km$^2$, the SKA is envisaged to provide unsurpassed sensitivity and leverage technological advances to implement a complex receiver to provide a large field of view through multiple beams on the sky. Many pathfinders and precursor aperture array telescopes for the SKA, operating in the frequency range of 10-300~MHz, have been constructed and operationalized to obtain valuable feedback on scientific, instrumental, and functional aspects. This review article looks explicitly into the progression of digital-receiver architecture from the Murchison Widefield Array (precursor) to the SKA1-Low. It highlights the technological advances in analog-to-digital converters (ADCs), field-programmable gate arrays (FPGAs), and central processing unit-graphics processing unit (CPU-GPU) hybrid platforms around which complex digital signal processing systems implement efficient channelizers, beamformers, and correlators. The article concludes with a preview of the design of a new generation signal processing platform based on radio frequency system-on-chip (RFSoC).
\end{abstract}

}]


\doinum{12.3456/s78910-011-012-3}
\artcitid{\#\#\#\#}
\volnum{000}
\year{0000}
\pgrange{1--}
\setcounter{page}{1}
\lp{1}

\section{Introduction}
\hfill

In the last few decades, there has been a renewed interest in astronomy at low radio frequencies ($\approx10 - 2000$~MHz) as this part of the electromagnetic spectrum is relatively unexplored. In the early decades of radio astronomy, an assemblage of connected antennas called aperture array telescopes was dominant until parabolic dishes began replacing them in the late 1960s \citep{garrett2013radio}. Since the onset of the 21${^{st}}$ century, advances in digital electronics, signal processing, computation, and storage technologies have enabled aperture array telescopes to make a comeback. While the requirement of high sensitivity, calibration techniques for wide fields of view, and mitigation of ionospheric effects are being addressed, human-made radio frequency interference (RFI) is one of the major obstacles in observing at low radio frequencies. RFI is a major challenge as a telescope shares the spectrum with many communication systems.

Radio telescopes operating at centimeter and meter wavelengths like the Giant Metrewave Radio Telescope \citep{gupta} have relied on an array of traditional dish antennas to sample a wide range of spatial frequencies. While the increasing spacing between interferometric dishes (longer baselines) helps achieve higher angular resolution, this might reduce surface brightness sensitivity for relevant angular scales. The surface-brightness sensitivity can be significantly improved with a large number of shorter baselines. However, there is a limit to the number of fully steerable large parabolic dishes that can be packed within a core area. Aperture array is preferable as it can provide a whole range of angular scales by having a configuration in which large groupings of antennas are contained within a compact area of a few kilometers diameter and the rest of the antennas spread out to large distances. An aperture array provides a large field of view, the capability to steer a beam electronically, flexibility in the number of beams and bandwidth, reliability, cost-effectiveness, and performance, as there are no moving parts. For a comparable collecting area afforded by an array of dish antennas, the large field of view of small antenna elements enables the collection of astronomical imaging information from a larger region of the sky in a given time. As low-frequency aperture array telescopes like the LOw-Frequency ARray \citep{van2013lofar}, Murchison Widefield Array \citep{lonsdale2009murchison}, and the Precision Array to Probe the Epoch of Re-ionization \citep{parsons2010precision} telescopes were coming online, radio astronomers began definitive plans to construct a large radio telescope called the Square Kilometre Array.
  
The SKA project is an international effort to build the world’s largest radio telescope \citep{dewdney2009square}. The SKA is planned to be built in phases, with SKA1 being the first. The first phase will consist of two telescopes, SKA1-Low and SKA1-Mid, covering the frequency range from 50~MHz to 15~GHz (aiming to reach 24~GHz). While the SKA1-Mid, a dish array, will be built in South Africa, the SKA1-Low, a low-frequency dipole array, will come up in Western Australia. SKA1-Low covers the lowest frequency band ranging from 50 to 350~MHz. SKA1-Mid aims to operate in the frequency range of 350~MHz to 24~GHz \citep{caputa2022ska}. The SKA, with its unprecedented sensitivity to detect signals from extremely weak radio sources, is expected to be a transformational instrument. In tune with the large-N (number of antennas) and small-D (area) architecture of modern low-frequency telescopes, SKA1-Low will consist of a large number of fixed antennas to provide an eventual collecting area of 1 km$^2$. 

The Murchison Widefield Array (MWA) radio telescope, located in outback Western Australia, is a low-frequency aperture array radio telescope operating in the frequency range between 80 and 300~MHz. MWA is the first precursor telescope for the SKA to become operational. A SKA precursor is a science and technology demonstrator located at one of the two sites shortlisted for the SKA. The Giant Metrewave Radio Telescope (GMRT), located in the western part of India, around 150 km east of Mumbai, is a major international facility for work in low-frequency radio astronomy. It consists of 30 fully steerable antennas of 45 meters in diameter that provide a total collecting area of $30,000$~m$^2$, covering a frequency range of 150~MHz to 1.5~GHz. In 2018 GMRT was granted the status of SKA Pathfinder -- a SKA-related technology, science, and operations demonstrator, but not located at one of the two sites shortlisted for the SKA.
 
This article reviews the progression of digital-receiver architecture from MWA to SKA1-Low. A description of the technology and implementation of the digital receiver to process signals from MWA is presented in Section \ref{sec:MWA}. Features of digital signal processing firmware implemented in the MWA digital receiver are reviewed. Section \ref{sec:uGMRT} gives details of the digital back-end for the upgraded GMRT. Section \ref{sec:SKA_Introduction} describes the receiver hardware architecture and firmware details of the digital receiver planned for SKA1-Low. The design of a new generation digital signal processing platform for radio astronomy that incorporates RFSoCs to eliminate discrete data converters is presented in Section \ref{sec:IPB}
 
\section{Murchison Widefield Array}
\label{sec:MWA}
The MWA is a low-frequency aperture array synthesis radio telescope operating in the frequency range 80 to 300~MHz \citep{lonsdale2009murchison}. The MWA was developed by an international collaboration, including partners from Australia, India, New Zealand, and the United States. It is located at the Murchison Radio-astronomy Observatory (MRO) in the Murchison Shire of Western Australia, which is $\approx800$~km north of Perth. The four key science goals of MWA are a) investigations of the Epoch of Reionization (EoR) power spectrum, b) Galactic and Extragalactic continuum and polarimetric studies, c) detection of transient sources, and d) solar and ionospheric science. MWA is the first of the four official precursors to the SKA to be completed. It is expected to provide valuable information related to the design of the SKA1-Low telescope's scientific, instrumental, and operational aspects. The Australian Square Kilometre Array Pathfinder \citep{schinckel2012australian}, located at the MRO, along with Hydrogen Epoch of Reionization Array \citep{deboer2017hydrogen} and MeerKAT \citep{manley2012meerkat} located in South Africa, are the other three precursor telescopes.

MWA \citep{tingay2013murchison} consists of 128-element aperture arrays (hereafter called tiles) spread over $\approx{}3$~km in diameter. Each tile comprises 16 antenna elements arranged in a square (4x4) configuration on a 5~m x 5~m wire mesh ground screen. In the 100~m diameter core area, about 50 uniformly spread tiles provided the necessary short baselines for EoR science; adjacent to the core are 62 tiles spread within a circle of 1.5~km diameter. The remaining 16 tiles are laid out within a 3~km circle to provide the longest baselines (highest angular resolution) required for solar imaging. At 150~MHz, all 128 tiles provide a total collecting area of 2752~m$^2$. In this array configuration, baselines range from 7~m to 2.8~km.

The primary antenna element is a dual-polarization crossed vertical bowtie. The polyvinyl chloride (PVC) hub to which the two aluminium arms of each bowtie get mounted also contains dual-polarization low-noise amplifiers (LNAs). Analog signals from all 16 dual-polarization antennas are combined in an analog beamformer. Each of the 32 analog paths of the beamformer contains an independent digitally controlled 32-step delay ranging from 0 to 13.5~ns in multiples of 435~ps. The outputs from all 16 delay stages are summed in a passive combiner to produce two tile beams on the sky, corresponding to each polarization. The beamformer outputs are transmitted over coaxial cables up to a maximum distance of 500~m to a station that hosts the MWA digital receiver.

A total of 16 digital-receiver units are deployed in the field to digitize (at 655.36 mega samples per second) and process dual-polarization signals from all 128 tiles. Although the sampled bandwidth from each tile is 327.68~MHz, output from digital-receiver units is limited to 30.72~MHz for further processing at the central processing facility located about 5~km from the core. A processed bandwidth of 30.72~MHz was deemed sufficient based on data distribution cost, the complexity of downstream real-time processing units, and continuum sensitivity. For 128 tiles with dual-polarization signals, 32768 signal pairs are multiplied in a correlator.

In the Phase 2 upgrade of MWA, 128 new tiles were added to the array configuration \citep{wayth2018phase}. Two sets of 36 tiles each were added to the existing core, and 56 tiles were placed beyond the array of Phase 1. The maximum baseline has been enhanced to $\approx5.3$~km. Only 128 tiles are operated at any given time as the digital receivers and correlator from Phase 1 were retained. The additional 128 tiles enhance the sensitivity of the instrument for EoR power spectrum measurement and improve the imaging capability of the array through enhanced angular resolution afforded by longer baselines, \textit{uv} coverage, and reduced confusion limit.

\subsection{Architecture of the Digital Receiver for MWA}
\label{sec:MWA_ArchDescription}
In the MWA receiver chain, instead of transporting beamformed analog signals from all 128 tiles to a central place that houses the digital receivers, it was advantageous to digitize and carry out first-stage digital signal processing in digital-receiver units in-field. Signals from 8 tiles are grouped and passed to a digital-receiver unit to optimize the number of such receiver units and their distribution based on the array configuration and station logistics. RF signals from 8 analog beamformers undergo signal level adjustments, power equalization, and additional anti-aliasing filtering in each receiver node's analog signal conditioning (ASC) board. All eight pairs of RF signals are bandpass filtered to limit their frequency content from 80 to 300~MHz and then passed to the digital-receiver unit. The digital receiver \citep{prabu2015digital} contains two analog-to-digital and filter bank (ADFB) boards, each containing four dual-channel 8-bit ADCs, AT84AD001C from e2v technologies, followed by four Virtex-4 SX35 FPGAs from Xilinx. The Virtex-4 family of FPGAs was introduced in 2004 on a 90~nm process technology. The ADCs operate at a clock rate of 655.36 mega samples per second (MSPs) and digitize a pair of RF signals corresponding to a tile. Two data streams from an ADC, each 8-bit wide, are passed to the corresponding Virtex-4 FPGA. In the FPGA, two instantiations of a 512-channel polyphase filter bank (PFB) split the sampled bands into 256 sub-bands, each 1.28~MHz wide.

\begin{figure*}[!ht]
     \centering\includegraphics[height= 10cm,width= 16cm]{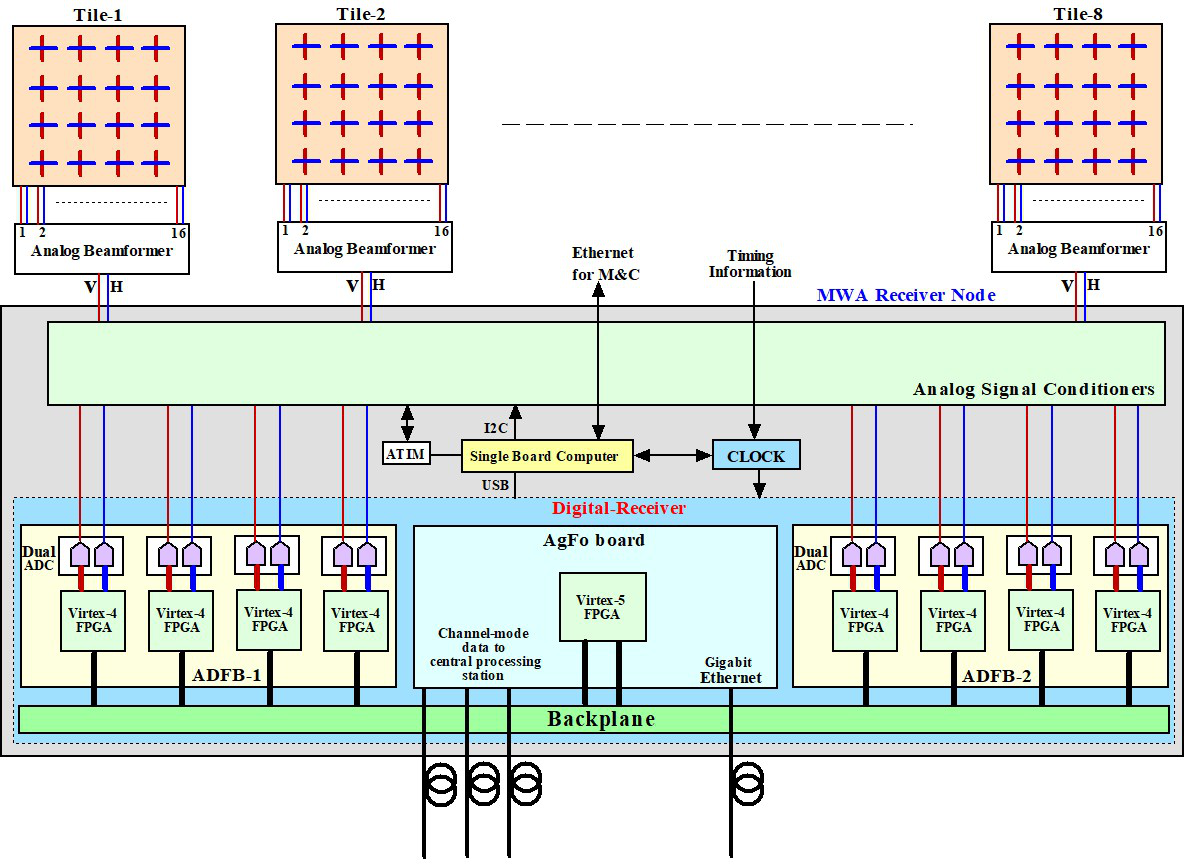}
         \caption{Each MWA receiver node consists of a digital receiver, analog signal conditioner, clock generation and distribution module, and a single board computer for M\&C. In the digital receiver, a custom-designed backplane provides input power and data interface between ADFB and AgFo.}
         \label{fig:MWA_BD}
\end{figure*}

In addition to the PFB channelizer, each Virtex-4 FPGA implements logic to select a user-defined subset of 24 PFB channels, corresponding to a bandwidth of 30.72~MHz. The 24 complex samples from the PFB stage are represented with a bit resolution of 5+5 bits to limit the data transport bandwidth. A data aggregator-formatter (AgFo) board, built around a Virtex-5 SX50T FPGA from Xilinx, contains four small form-factor pluggable (SFP) electrical-to-optical fiber conversion modules for high-speed data transfer. The Virtex-5 family of FPGAs was introduced in 2006 on a 65~nm process technology. The Virtex-5 FPGA on the AgFo board implements logic to gather PFB data corresponding to 8 tiles streaming out from 2 ADFBs and performs reordering and formatting of data. 3 SFP modules route serialized data (in an 8b/10b encoded format) via optical fiber links to the central processing station for further signal processing and correlation. The fourth SFP module provides a Gigabit Ethernet interface for diagnostics. A custom-designed backplane provides the interface between ADFB boards and the AgFo board. In addition to routing PFB data from ADFB boards to AgFo, the backplane distributes power, clock, and synchronization signals. A test port is provided on the backplane to capture diagnostic data. A single-board computer running an embedded Linux operating system controls receiver node functions like a) analog beamformer settings, b) monitoring ASC board, c) streaming configuration bitstream of all FPGAs via a USB interface, d) setting modes of operation, and e) thermal and power management. Figure~\ref{fig:MWA_BD} shows a block diagram of the MWA receiver node.

A summary of sub-components contained in each receiver node \citep{prabu2015digital}:
\begin{itemize}
\item Analog Signal Conditioner(ASC)
\item Antenna interface module (ATIM) to control the tile pointing
\item ADFB boards for digitization and channelization
\item AgFo board for data gathering and formatting
\item Single board computer (SBC) for monitor and control (M\&C) interface
\item Regulated power supply for sub-components, tile beamformer	
\item Clock module to generate sampling clock, processing clock and synchronization pulse at one-second interval
\end{itemize}

As the receiver node is deployed in-field and exposed to a wide range of ambient temperature from about 0$^\circ$ to 50$^\circ$~C, the receiver electronics is housed in a weather-proof RF-shielded enclosure. The RF-shielded enclosure prevents RFI emission from electronic modules from interfering with the normal operation of the antennas and meets the demanding radio-quiet requirements of the MRO. A refrigeration unit maintains a stable operating temperature of $\approx25^\circ$~C. Sixteen such receiver nodes are deployed for the 128-tile MWA system. All 16 digital-receiver units for the MWA were developed at Raman Research Institute, Bengaluru.

\subsection{Signal Processing in the MWA Digital Receiver}
\label{sec:MWA_DSP}

By operating the digital receiver in a specific mode of operation (refer to Section \ref{sec:MWA_ModesOfOperation}) or by utilizing meta-data functionality in the normal mode of operation, a remote station can assess the power of the analog signal at an ADC's analog input port \citep{prabu2015digital}. In the channel mode of operation of the digital receiver, which is its normal mode of operation, the M\&C interface can be used to obtain ADC output power, RFI scenario, and power spectrum at the output of the channelizer. Based on the assessment of signal power using one of the above methods, the amplification setting in the ASC unit is adjusted to optimize the signal power.

After digitizing the analog signal, frequency channelization is a crucial digital signal processing step employed in modern radio telescopes to split a sampled band into narrow sub-bands. Fast Fourier Transform (FFT) as a filter bank suffers from two drawbacks: poor side-lobe performance and scalloping loss. A PFB is a computationally efficient implementation of channelization. The design of the prototype filter bank preceding the FFT block offers great scope to control the flatness of sub-bands and provides excellent suppression of out-of-band signals \citep{bellanger1974tdm}. In the MWA digital receiver, ADC data at 655.36~MSps is streamed to a 512-channel coarse (first-stage) PFB. As the input time series to the coarse PFB is real-valued, it splits the sampled band of 327.68~MHz into 256 sub-bands, each 1.28~MHz wide. The coarse PFB is a "critically-sampled" PFB as the separation between adjacent spectral channels equals the channel Nyquist bandwidth of 1.28~MHz \citep{morrison2020performance}. With eight filter taps per channel, the length of the prototype filter is (512x8) =4096 (coefficients). In the PFB implementation, the bit precision of filter coefficients and the twiddle factors of the FFT stage is 12 bits. The native bit precision of complex 1.28~MHz spectral channels is 16+16 bits. Measurements of PFB channel response have shown that the pass-band flatness is within $\pm$0.1~dB, and the adjacent channel suppression is better than 50~dB. The bit resolution of complex PFB output channels is limited to 5+5 bits to minimize the data transport bandwidth. The voltage gain section at the PFB output is used to equalize the pass-band gain variations and maintain the bit-occupancy to represent the sky signal, with adequate headroom for RFI. The central processing station implements a 128-channel critically-sampled fine PFB stage in four dedicated FPGA-based PFB boards \citep{tingay2013murchison}. Each of the 1.28~MHz wide spectral channels of the coarse PFB is split into 128 sub-bands to obtain a spectral resolution of 10~kHz.

In a polyphase filter bank, the transition band of a channel refers to the frequency response between the pass-band and the stop-band. The width of the transition band, determined by the design of the prototype filter, permits a trade-off between the adjacent channel leakage performance and maximizing the fraction of uncorrupted pass-band \citep{morrison2020performance}. For radio astronomy applications, as maximizing the fraction of uncorrupted pass-band (within the Nyquist bandwidth of a critically-sampled PFB channel) is preferred, there will be aliasing at either edge of a coarse PFB channel. When a fine PFB stage further channelizes a coarse PFB channel for finer spectral resolution, spectral channels at either end of the fine PFB output may have to be discarded as they are corrupted due to aliasing. Discarding spectral channels results in periodic discontinuities across the sampled band. However, if the first-stage PFB is an oversampled PFB, the Nyquist bandwidth is wider than the separation between channels \citep{harris2003digital}. The wider Nyquist bandwidth, whose width is decided by the oversampling factor, permits extraction of the alias-free portion of the band for further processing. A contiguous band without periodic discontinuities is obtained by stitching together alias-free spectral channels at the output of fine PFB.

\vspace{1em}
\begin{table}[htb]
	\tabularfont
	\caption{Specifications of MWA digital receiver}
	\begin{tabular}{p{1.5in}p{1.5in}}
		\topline
		Digital receiver parameter&Value\\\midline
		Number of analog input signals & 8 dual-polarization signals\\
		Frequency range of analog signal & 80-300~MHz\\
		ADC used & 8-bit AT84AD001C from e2v technologies\\
		ADC sampling clock&655.36~MHz\\
		Sampled bandwidth&327.68~MHz\\
		Processed bandwidth&30.72~MHz\\
		FPGA used in ADFB&XCV4SX35-10FFG668 Virtex-4 from Xilinx\\
		FPGA used in AgFo&XC5VSX50T-10FFG665C Virtex-5 from Xilinx\\
		Channelizer (first-stage)&512-channel critically-sampled PFB\\
		Number of PFB output channels & 256\\
		Spectral resolution&1.28~MHz\\
		Length of prototype filter & 4096 filter coefficients\\
		Bit resolution&12-bit representation of filter coefficients, FFT twiddle factor\\
		Passband ripple& $\pm0.1$~dB\\
		Sidelobe performance&better than 50~dB\\
		PFB output bit resolution&5+5 bits\\
		Output data rate & 5.4~Gbps on 3 optical fibers\\	
		\hline
	\end{tabular}
	\label{table:MWA_Spec}
\end{table}

Table~\ref{table:MWA_Spec} lists the major specifications of the MWA digital receiver. FPGA firmware for the MWA digital receiver was designed using the Very high-speed integrated circuit Hardware Description Language (VHDL). While Xilinx's Integrated System Development (ISE) software tool was used for the synthesis and analysis of HDL designs, Mentor Graphics' ModelSim was used for the simulation of HDL designs.

\vspace{2em}
\subsection{Modes of Operation of the Digital Receiver}
\label{sec:MWA_ModesOfOperation}

Telescope performance can be affected if the performance of any of its sub-systems is deficient. Hence, effective fault detection and diagnostic methods are vital to improving sub-systems' performance and availability. It is more so in the case of MWA digital receivers as they are distributed across the array, necessitating remote monitoring and control. Apart from the channel mode of operation, which produces science-quality data, the digital receiver can be programmed to operate in various modes like Burst mode, Raw-data mode, and Diagnostic mode \citep{prabu2015digital}. In the channel mode of operation, the digital receiver continuously outputs a user-defined subset of 24 PFB channels (of a tile) to the AgFo board. The AgFo board aggregates data from all eight tiles; aggregated data is reordered and formatted into data packets that are streamed out of the AgFo board on three optical fiber cables. In this mode, the AgFo board also has a provision for routing any one channel, out of 24 channels, from all eight tiles via the Gigabit Ethernet interface for continuous monitoring. Additionally, a station beamformer mode of operation forms a voltage beam from all eight tiles connected to a digital receiver\citep {prabu2014full}.

The burst mode of operation permits one complete PFB output frame consisting of 256 spectral channels in its native bit precision of 16+16 bits to be streamed out via the Gigabit Ethernet interface. In this mode, full precision data corresponding to 327.68~MHz bandwidth from all eight tiles connected to a digital receiver is streamed once every 1024 PFB frames. As full bandwidth data is preferred over continuous-time data, it is streamed once every 800~$\mu$s. In raw data mode, 256 ADC samples in their native 8-bit precision are sniffed once every 800~$\mu$s and routed via the Gigabit Ethernet interface. The diagnostic mode is useful when the receivers need testing and debugging. In this mode, there is a provision to insert test patterns at various points in the receiver chain to aid in systematic testing and fault isolation.

\section{Upgraded Giant Metrewave Radio Telescope}
\label{sec:uGMRT}

GMRT consists of an array of 30 antennas, each of 45~m diameter, spread over a region of 25~km extent, and operating at five different wavebands from 150 to 1450~MHz. For the legacy system \citep{roy2010real}, the maximum instantaneous operating bandwidth at any frequency band is 32~MHz. GMRT underwent a major upgrade \citep{gupta2014gmrt}, intending to provide near-seamless frequency coverage over 120 to 1500~MHz with improved feeds and receivers, along with a maximum instantaneous bandwidth of 400~MHz. The four sets of feeds and receivers that cover this range have the RF signals transmitted over optical fiber to the central receiver building, where they are digitized and processed in the GMRT wideband back-end (refer to Section \ref{sec:uGMRT_GWB_Introduction}). In addition to the increase in the maximum instantaneous bandwidth by a factor of 10, the other major improvements in the upgraded GMRT (uGMRT) compared to the legacy GMRT system are a) the larger number of spectral channels (from 256–512 to 2048–16384), b) polyphase filtering to reduce spectral leakage, c) the increase in the number of beams (from two to four), d) options for real-time RFI filtering \citep{buch2016towards}, e) Walsh demodulation capability, and f) a real-time coherent dedispersion \citep{de2016real} system that supports the much larger bandwidth.

\subsection{GMRT Wideband Back-end (GWB)}
\label{sec:uGMRT_GWB_Introduction}

Historically, interferometer back-ends have been hardware-based, using dedicated Application-Specific Integrated Circuits (ASICs) and FPGAs \citep{perley2009expanded}. With the increasing availability of high-performance computing systems, there has been a shift towards software-based alternatives \citep{roy2010real, deller2007difx}. Compared to hardware back-ends, some advantages of such systems are easier off-the-shelf availability, relatively shorter development times, higher flexibility, and easy upgradability. Graphics Processing Units (GPUs) have ushered in a new radio astronomy digital back-ends era. These massively parallelized, compute-intensive systems prove highly suited for developing large-N, large bandwidth correlators. Additionally, hybrid correlators, which distribute the stages of the correlation process onto different types of compute engines, have proved remarkably useful for such large-N, large bandwidth correlators. One such hybrid back-end system is the GMRT Wideband Back-end \citep{reddy2017wideband} developed for the upgraded GMRT using FPGAs and GPUs.

\subsubsection{Design of the GWB}
\label{sec:uGMRT_GWB_Design}
\begin{figure*}
	\centering\includegraphics[height= 9.5cm,width= 16cm]{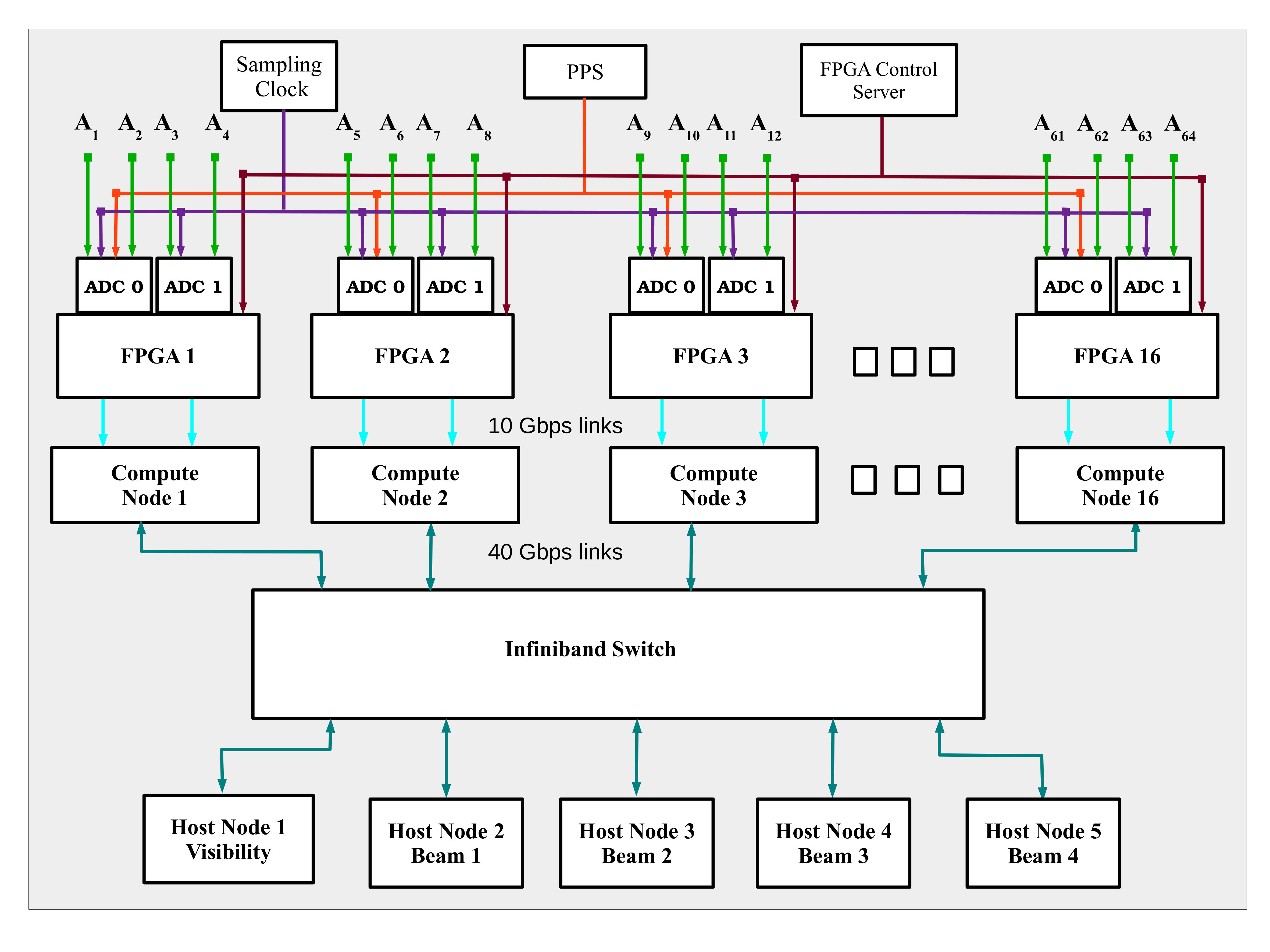}
	\caption{Top-level design of the GWB (from Fig. 1 of \citep{reddy2017wideband}): A1, A2, . . ., A64 are Baseband converted analog signals from antenna. GWB uses a hybrid design of FPGA, CPU and GPU, with an InfiniBand interconnect for data re-distribution between the compute nodes. Visibility output and beam outputs are collected in the Host nodes using the same InfiniBand network and the outputs are recorded onto the disks. PPS is the Pulse Per Second signal for synchronization.}
	\label{fig:TopLevelDesign}
\end{figure*}

The GMRT Wideband Back-end can support a maximum of 400~MHz instantaneous bandwidth from 64 input signals (dual-polarization inputs from 32 antennas) for interferometry and beamformer modes. GWB also supports lower bandwidths from 200~MHz to 1.5625~MHz. The design combines an FPGA-based front-end with a GPU-based software compute workhorse, an InfiniBand-based network, and optimized software to achieve the target compute and I/O requirements. The top-level design of the GWB is illustrated in Figure~\ref{fig:TopLevelDesign}. The FPGA boards, connected to high-speed ADCs, perform the digitization and packetization of the voltage signals from all the antennas, while off-the-shelf compute node servers fitted with GPUs acquire the data, compute visibility products, and beamformer outputs, which are then recorded on hard disks for post-processing and analysis.

The implementation is a time-slicing model, where each compute node gets a slice of contiguous time-series data from all the antennas, on which it performs spectral conversion (F), correlation (X), and Incoherent Array (IA) and/or Phased Array (PA) beam formation operations. The data is digitally down-converted for narrowband mode observations and decimated before the spectral conversion stage. Dedicated 10 Gigabit Ethernet links transfer digitized data from the FPGAs to the compute nodes. A 40~Gbps InfiniBand network is used to redistribute data for time slicing and gathering results onto the host nodes for the final recording \citep{reddy2017wideband}.

The time-slicing model used here has several features that make it attractive for a software/hybrid design compared to a frequency-slicing model (where the F stage is usually implemented on the FPGA). Before the F stage, the bulk time delay correction for different antenna signals has to be carried out. This is much more conveniently done for the time-slicing model, especially for longer baselines (few kilometers) at high sampling rates, as the buffering of digitized data for delay correction is performed on CPU memory, which is generally a few gigabytes, compared to much smaller memory size on the FPGA. Keeping the F and X stages together on the GPU (as done in GWB’s time-slicing model) allows for optimal, tightly coupled code. At the same time, it greatly simplifies the FPGA design by freeing up precious resources that are utilized for the implementation of signal quality enhancing features like RFI filtering and Walsh demodulation.

\subsubsection{Compute and I/O requirements}
\label{sec:uGMRT_GWB_ComputeIO}

For the GWB, the computation requirements come to 2.9~Tflops for FFT (16384 points), 6.8~Tflops for Multiplication and Accumulation, and 0.1~Tflops for phase-shifting operation for the residual delay and fringe correction. For the beamformer, assuming two IA beams and two PA beams, the compute load comes out to be 0.45~Tflops. Overall, the computation requirement for the GWB is around 10.2~Tflops. The maximum overall input data rate to be handled is 25.6~GB/s considering 32 antennas (rounding the number of antennas to the nearest power of two), 400~MHz bandwidth, and 4 bits per digitized sample. The output visibility data rate is about 51~MB/s for 2048 spectral channels, 0.67~s integration time, and floating-point storage data type. The raw output beam data rate is 209~MB/s for IA or PA beamformers considering 2048 spectral channels, the integration time of $40~\mu$s, and the short integer storage data type.

\subsubsection{Implementation}
\label{sec:uGMRT_GWB_Implementation}

The design uses Xilinx Virtex-5 FPGA-based Reconfigurable Open Architecture Computing Hardware (ROACH) boards developed by the CASPER group, 8-bit ATMEL/e2v-based iADC developed by the CASPER group, DELL T630 servers as compute and host nodes, and NVIDIA Tesla K40c GPUs. Myricom 10GbE CX4 network interface cards (NICs) and cables are used for data transfer between FPGAs and compute nodes. A Mellanox QDR InfiniBand switch and Mellanox InfiniBand NICs are used for data redistribution between the compute nodes and gathering visibility and beam data onto the host nodes.

Each iADC can sample two analog input signals, and each ROACH board has a provision for handling two iADCs or digitized data from four input signals. The data from each iADC, i.e., for two input signals, is bundled into a User Datagram Protocol (UDP) packet that can be sent through one of four 10GbE CX4 ports on the ROACH board. Thus, four input signals can be sampled, packetized, and sent over two 10GbE links, to be received by a single, dual-port 10GbE NIC on a PCIe x8 or x16 slot of a compute node (Dell T630 server). The data thus received is redistributed between the compute nodes over the 40~Gbps InfiniBand interconnect. The redistributed data, representing a slice of contiguous time series data from each antenna, is passed onto NVIDIA GPUs for further processing. Each compute node can handle two GPUs, with each GPU processing half the slice of the contiguous time series data received by the compute node. Overall, the GWB has been implemented using 32 iADCs, 16 ROACH boards, 16 Dell T630s as compute nodes, 1 Dell T630 as the visibility host node and 4 Dell T620/T630s as the beam host nodes, 32 NVIDIA Tesla K40c GPU cards, and a 32-port Mellanox InfiniBand QDR switch.

Table~\ref{table:uGMRT_Rx_Specs} lists the major specifications of the wideband back-end for uGMRT.
\begin{table}[htb]
	\tabularfont
	\caption{Specifications of the wideband backend for uGMRT}
	\begin{tabular}{p{1.5in}p{1.55in}} 
		\topline
		Number of analog input signals & 32 dual-polarization signals\\
		ADC used & 8-bit, dual-channel AT84AD001B from e2v technologies\\
		ADC sampling clock & 800~MHz\\
		Sampled bandwidth & up to 400~MHz (max)\\
		Processed bandwidth & 100, 200 or 400~MHz\\
		Bits per Sample&8 for 100~MHz, 4 for 400~MHz\\
		FPGA used in ROACH board & XC5VSX95T Virtex-5 from Xilinx\\
		Channelizer & Single-stage critically-sampled PFB\\
		Number of PFB output channels & Configurable from 2048- to 16384-points\\
		Full Stokes capability&Yes\\
		Walsh Demodulation&Yes\\
		RFI filtering&Yes\\
		Minimum visibility integration time&671~milliseconds\\
		Coarse and fine delay tracking range&$\pm128$~microseconds\\
		Fringe rotation range&up to 5~Hz\\
		Number of sub-arrays&4\\
		Number of beams&4 (Incoherent array/Phased array)\\
		\hline
	\end{tabular}
	\label{table:uGMRT_Rx_Specs}
\end{table}

\subsection{Ongoing developments}
\label{sec:uGMRT_OngoingDevlopements}

The FPGA-CPU-GPU design enables the addition of more features to the system. Apart from adding features like Walsh demodulation and RFI filtering, work is ongoing to develop a back-end parallel to the GWB to have raw data recording and/or beamforming with multiple beams. For this work, the two 10GbE ports on the ROACH board (out of four) can be used to send a copy of sampled data to the parallel system. For the raw voltage recording, testing on NVMe (non-volatile memory express) disks is ongoing, which can record up to 2~GB/s. NVMe disks are also being used to record voltage beam data coming to the beam hosts. Work is also ongoing to extract multiple narrow bands from the wideband by implementing a polyphase filter bank on GPU.

\section{Square Kilometre Array-Low}
\label{sec:SKA_Introduction}

The Square Kilometre Array project is an international endeavour to build the world’s largest radio telescope. By constructing thousands of dish antennas and up to a million low-frequency wire antennas, SKA will eventually provide a collecting area of 1~km${^2}$, enabling astronomers to probe the universe with unprecedented sensitivity, angular resolution, and sky coverage. Fourteen member countries and a number of organizations are participating in the design and development of the SKA. SKA1-Mid will be made up of about 200 dishes in the Karoo region of central South Africa, and the SKA1-Low telescope will be built using $2^{17}$ low-frequency dual-polarization wire antennas at the MRO. Apart from radio quietness and characteristics of the atmosphere above the two sites, major criteria like physical characteristics of the site, ability to provide connectivity across the vast expanse of the telescope, cost of infrastructure, operation, and maintenance were crucial in the choice of the sites. Key science projects of the SKA are: Probing the Dark Ages; Galaxy evolution, cosmology, and dark energy; Origin and evolution of cosmic magnetism; The cradle of life - searching for life and planets; and Strong-field tests of general relativity with pulsars and black holes \citep{schilizzi2008square}.

The Low-Frequency Aperture Array (LFAA) covers the lowest frequency band of the SKA ranging in frequency from 50 to 350~MHz. The LFAA refers to that part of the SKA1-Low telescope containing antennas, amplifiers connected to the antenna terminals, and the local processing of signals that logically groups the antennas into stations to produce station beams, facility for control, monitoring, and calibration. To meet the sensitivity requirements of key science projects of SKA1-Low, LFAA will require a large collecting area. In this frequency range, a large collecting area can be realized by employing a considerable number of phased array elements. 131072 dual-polarization antennas of LFAA are arranged as 512 stations. Two hundred fifty-six antennas placed in randomized positions within an effective diameter of $\approx38$~m form a station. Antennas within a station are logically grouped as 16 tiles, each consisting of 16 elements. The antenna configuration of SKA1-Low is such that 75\% of the antennas are arranged within a dense core of about 2 km diameter, and the rest spread out in three quasi-spiral arms providing a maximum baseline of about 65~km. One of the primary aims of SKA1-Low is to achieve high-surface brightness sensitivity for a wide range of angular scales as it aids in the detection of Epoch of Reionization. As over 75\% of the baselines are short, the surface-brightness sensitivity will improve significantly. In Phase 1, the total collecting area of SKA-Low will be about 0.4~km${^2}$.

The primary antenna element is a dual-polarization log-periodic dipole (LPD) array fixed to a metallic wire mesh ground screen. The ground screen helps insulate antennas from soil conditions and aims to provide a similar environment for all antennas \citep{de2015skala}. Across the operating frequency range, the LPDs are noise-matched to realize the optimal performance of the low-noise amplifiers connected to the antenna elements. RF signals amplified by LNAs and band-limited to 50-350~MHz by appropriate filters are converted to optical signals and transmitted as RF-over-Fiber (RFoF). RFoF links carrying RF signals from individual log-periodic dipoles are received at the central processing facility (CPF) in the field. At the facility, the Italian Tile Processing Module (TPM) and sub-systems that support and coordinate TPM activities form the signal processing sub-system (SPS) of the LFAA. Each TPM digitizes dual-polarized RF signals from 16 antennas, channelizes the sampled bands into narrow spectral channels, and generates a partial station beam (tile beam). Partial station beams from 16 such TPMs are summed to form a station beam. Station beams are transported over optical fiber links to the Central Signal Processor (CSP) for further processing. The Monitor, Control, and Calibration Sub-system (MCCS), consisting of a number of high-performance servers and high-speed switches, implements the local monitoring, control, and calibration functions for the stations and associated sub-systems. The MCCS interfaces with TPMs and supporting sub-systems within the SPS.

\subsection{Architecture of the Tile Processing Module}
\label{sec:SKA_TPM_Arch}

\begin{figure*} 
	\centering\includegraphics[height= 9.75cm,width= 16.0cm]{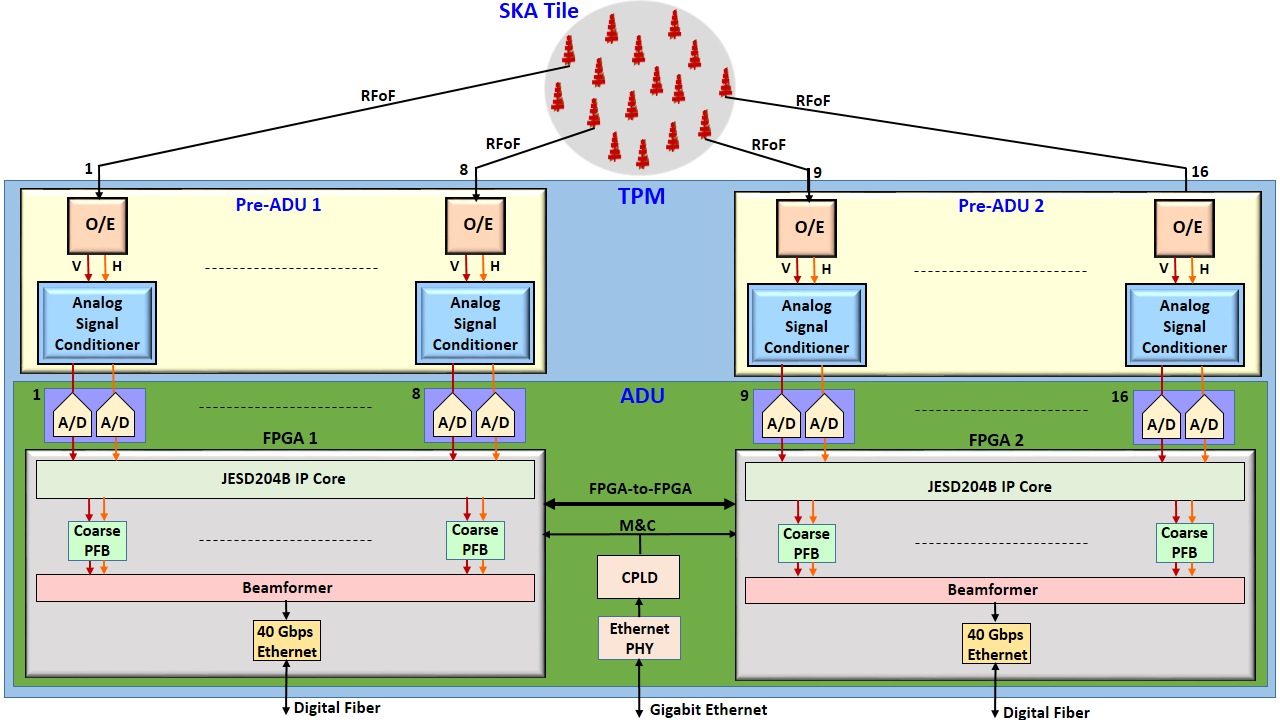}
	\caption{At the core of the Italian Tile Processing Module are two Pre-ADU boards and an ADU board. After optical-electrical conversion (O/E) and analog signal conditioning in the Pre-ADU boards, 16 dual-polarization signals from a SKA tile are digitized, channelized and beamformed in the two Kintex UltraScale+ FPGAs in the ADU board. Data flown through 16 such daisy chained TPMs produces a station beam.}
	\label{fig:ITPM_BD}
\end{figure*}

The signal processing platform for LFAA, the Italian Tile Processing Module, is built around two Pre-ADU boards and an Analog-to-Digital Unit (ADU) board \citep{schilliro2020design}. The Pre-ADU boards receive 16 RFoF links carrying dual-polarization RF signals from 16 antennas. It converts them to electrical signals and implements RF signal conditioning like filtering, amplification, and equalization of all 32 (16 dual polarization) RF signals. The power level of the RF signal in the Pre-ADU is adjusted using a digital step attenuator controlled via the Serial Programming Interface (SPI) signals from the ADU board. The ADU board contains 16 dual-channel 14-bit ADCs, AD9695BCPZ from Analog Devices, for analog-to-digital conversion of 32 RF signals at 800~MSps. JESD204B (standardized serial data interface) data transmit block inside the ADC manages the serial data interface between ADC and Kintex UltraScale+ FPGA, XCKU9P from Xilinx, on the ADU. Two Kintex FPGAs, built using 16~nm process technology and introduced in 2015, form the primary processing element of the TPM. Each FPGA captures data streams from 8 ADCs for further processing. A complex programmable logic device (CPLD) on the ADU board controls the power-up sequence of onboard devices like FPGAs, ADCs, phase-locked loops for clock generation, and settings in the analog section. The Local Monitor and Control (LMC) sub-system connects to the ADU through a Gigabit Ethernet interface \citep{comoretto2017signal}. Figure~\ref{fig:ITPM_BD} shows a block diagram of the TPM. 

FPGAs, through the instantiation of JESD204B interface blocks, collect data from each ADC over two JESD204 data lanes \citep{naldi2017digital}, corrects for up to $\pm120$~m of cable length mismatches, implements a 1024-channel oversampled (by a factor of 32/27) coarse polyphase filter bank to split the 400~MHz sampled band into 512 spectral channels. The spacing between spectral channels is 0.781~MHz. A dynamic phase correction provided by the MCCS is applied to each PFB spectral channel to compensate for the time delay. Partial beams are formed within a TPM after applying instrumental gain, phase and polarization calibration, and weighting with appropriate beamforming coefficients. At the PFB output, it is possible to either select a set of spectral channels and generate multiple beams or select multiple regions of the output spectrum and form multiple beams using the chosen spectral regions. Sixteen such TPMs are required to process signals from all 16 tiles of a station. Using ADU's 40~Gbit Ethernet interface (QSFP+), multiple TPMs are connected through a high-speed data switch \citep{comoretto2017signal}. As the partial-beams travel through the daisy-chained TPMs, it gets summed up to form the station beam. The station beam is transferred via the same high-speed network to the CSP for fine channelization, array beamforming, and correlation.

In the normal mode of operation, the TPM provides a host of diagnostic and calibration features like a total power detector at the ADC output, generation of autocorrelation spectrum of an antenna, or cross-power spectrum for a pair of antennas for calibration purposes \citep{comoretto2017signal}. 

ADU boards, designed to interface with a backplane, are in a compact 6U form factor. A sub-rack system houses 8 TPMs, a management board, and a power supply unit \citep{schilliro2020design} interfaced with the backplane. In addition to input power distribution to the connected boards, the backplane provides each ADU with various interface links to manage TPM operations. The Management board generates the required functions to coordinate TPM operations and distributes clock and synchronization signals received from a central facility. A Gigabit Ethernet connection provides an interface to configure, control, and monitor TPMs. As part of the SPS, two such sub-racks are required to process signals from 256 antennas of a station. In all, 8192 TPMs are required to process signals from all 512 stations of the LFAA. Table~\ref{table:TilesStationsTPMs} provides a quick summary of the number of antennas, tiles, TPMs, and sub-racks required for an LFAA station.

\begin{table}[htb]
	\tabularfont
	\caption{LFAA station -- Summary of tiles, TPMs, and sub-racks}
	\begin{tabular}{p{1.75in}p{1.25in}}
		\topline
		Parameter&Value\\\midline
		Total LFAA antennas&131072 dual-pol.\\
		Number of antennas in a tile&16\\
		Number of tiles in a station&16\\
		Number of antennas in a station&256\\
		Number of stations&512\\
		Number of TPMs per tile&1\\
		Number of TPMs per station&16\\
		Number of TPMs for all 512 stations&8192\\
		TPMs housed in a sub-rack&8\\
		Sub-racks per station&2\\
		\hline
	\end{tabular}
	\label{table:TilesStationsTPMs}
\end{table}

\subsection{Signal Processing in the TPM}
\label{sec:SKA_TPM_DSP}

SKA1-Low, expected to become a transformational instrument due to its high sensitivity and wide frequency coverage, requires complex signal processing sub-systems to meet the challenging requirements of key science projects. In the LFAA receiver chain, a frequency domain approach is followed for digital beamforming and correlation (F-X correlator) to generate the visibilities. The TPM, which forms the core of the SPS installed at the central processing facility, implements the first digital signal processing stage for the LFAA. TPMs receive a 10~MHz reference frequency signal and a synchronization signal (Peak-Per-Second (PPS)) aligned to the Universal Time Coordinated from the synchronization and timing (SAT) system. The reference signal is used to generate the 800~MHz sampling clock to the AD9695 ADCs of TPM. Although AD9695 has 14 bits of resolution with an effective number of bits (ENOB) of ˜10.5, due to bandwidth limitation, only 8 bits per sample are sent to the FPGA. With this choice, as the ENOB is about 7.9, it nearly performs like an ideal 8-bit ADC \citep{comoretto2017signal}. The data rate from each dual-channel ADC is 12.8~Gbps.

\subsubsection{Channelization}
\label{sec:SKA_TPM_DSP_Channelization}

The LFAA implements a two-stage PFB. The first-stage channelizer implemented in the TPM is an oversampled PFB. Further down the signal processing chain, for applications requiring high spectral resolution, a critically-sampled PFB is realized in the CSP as the second-stage channelizer. The first-stage channelizer is a 1024-channel PFB with 14 filter taps per channel and operates at an oversampling factor of $32/27$. The output of the polyphase filter section is processed by an optimized 1024-point FFT block \citep{comoretto2020signal}. As the input to the PFB is a real-valued sequence, the 512 complex FFT output channels across the sampled band of 400~MHz result in spacing between channels of 0.781~MHz. Due to oversampling, the commutator, which feeds successive samples from the ADC to the polyphase filter section, advances by only 864 arms of the filter section instead of 1024 arms. One hundred sixty samples from the previous PFB frame are retained in generating a new output frame. An increase in the processing speed by the oversampling factor is required to ensure that ADC data samples are not lost. As a result, the Nyquist bandwidth of each channel of the first-stage PFB is 0.925~MHz. After further channelization in the second-stage PFB, only the region corresponding to 0.781~MHz, free from aliasing, is retained for further processing. The prototype filter design of the first-stage PFB results in a passband ripple of $\pm0.2$~dB. While the stop-band attenuation is better than 60~dB, across the spectrum, the average stop-band attenuation is better than 100~dB \citep{comoretto2017signal}. High stop-band attenuation constrains the deleterious effect of a strong RFI to a few channels and prevents the masking of weak signals across the band. At the output of first-stage PFB, with the bit resolution of each channel set to 12+12 bits, the aggregate data rate from all 16 channelizers is 274~Gbps \citep{schilliro2020design}.

\subsubsection{Beamforming}
\label{sec:SKA_TPM_DSP_Beamforming}

The process of beamforming requires aligning the phases of the incoming signal from antennas across the array. Although time-delay beamforming is best suited, realizing true-time delay using a set of switchable cable lengths or tracks on a printed circuit board (PCB) is cumbersome. Not only does realizing true-delay occupy considerable space, but it can also be limited in resolution and suffer from repeatability issues. Additionally, if multiple beams are required, it can be complicated and expensive to implement. Beamforming in the frequency-domain involves a) digitizing signals from each antenna, b) channelizing the sampled band into narrow sub-bands, and c) multiplying each sub-band with a complex function prior to the summation of corresponding sub-bands. Frequency domain beamforming offers flexibility in the beamforming process and aids in detecting and managing RFI-affected sub-bands. Apart from the required geometric phase to point the beam in a specific direction, the complex multiplication for each sub-band includes weights for the receiver (gain, phase), atmospheric, and polarization calibrations. Corresponding spectral channels from each antenna are summed to produce a beam. The sub-bands are multiplied with different complex functions to form multiple beams in the sky.

Post channelization, the LFAA implements digital beamforming in the frequency-domain. A subset of channelizer output samples from both polarization of all antennas are sent to MCCS for computation of the complex calibration coefficients and delay required for beam steering for each element \citep{schilliro2020design}. The complex correction values for each dual-polarized antenna are arranged as a (2x2) matrix. By phase referencing every antenna in a tile to a common antenna in a station, each TPM generates a partial beam (tile-beam). The high-speed FPGA-to-FPGA interface transfers the tile beam's odd and even spectral channels for further processing in respective FPGAs. In the next step, tile beams (16+16 bits) from TPMs corresponding to a station are propagated as travelling sum packets on the 40~Gbit Ethernet interfaced to a high-speed data switch. With the protocol overhead, the data rate of the travelling sum is 23.2~Gbps. The final TPM produces the station beam with a bit resolution of 8+8 bits. The station beam is streamed to the CSP at 11.6~Gbps for further processing. Due to limitations in the hardware, the product of the number of beams and the bandwidth of each beam is limited to 300~MHz. Up to 48 beams on the sky can be generated per SKA1-Low station. As the number of sub-bands is constrained to 8, the bandwidth of each of the 48 beams is 6.25~MHz \citep{comoretto2020signal}. Table~\ref{table:TPM_Specs} lists the major specifications of the LFAA digital receiver.

 \begin{table}[htb]
	\tabularfont
	\caption{Specifications of the TPM used in LFAA}
	\begin{tabular}{p{1.5in}p{1.55in}}
		\topline
		TPM parameter&Value\\\midline
		Number of analog input signals & 16 dual-polarization signals\\
		Frequency range of analog signal & 50-350~MHz\\
		ADC used in ADU&14-bit AD9695BCPZ from Analog Devices\\Sampling clock&800~MHz\\
		Sampled bandwidth&400~MHz\\
		Processed bandwidth&300~MHz\\
		FPGA used in ADU&XCKU9P Kintex UltraScale+ from Xilinx\\
		Channelizer (first-stage) & 1024-channel oversampled (32/27) PFB\\
		Number of PFB output channels & 512\\
		Filter taps per channel & 14\\
		Bit resolution of filter coefficients & 18-bit\\
		Spacing between channels&0.781~MHz\\
		Passband ripple& $\pm0.2$~dB\\
		Sidelobe performance&better than 60~dB\\
		PFB output bit resolution&12+12 bits\\
		Partial-beam bit resolution&16+16 bits\\		
		Output data rate from ADU&23.2~Gbps on 40~Gbit Ethernet interface\\
		\hline
	\end{tabular}
	\label{table:TPM_Specs}
\end{table}

FPGA firmware for the TPM was designed using VHDL. Vivado Design Suite from Xilinx was used for the synthesis and analysis of HDL designs.

\section{New generation receiver architecture for a low-frequency radio telescope}
\label{sec:IPB}

Since the introduction of FPGA more than three decades ago, its architecture has constantly evolved to meet challenging applications. In 2011, Xilinx introduced the Zynq architecture in which the traditional FPGA logic was integrated with on-chip processors, alleviating the challenges in interfacing user-programmable logic with external processors. In 2017, the integration of high-speed data converters into All Programmable multiprocessor systems-on-chip (MPSoC) architecture resulted in the first-generation Zynq UltraScale+ RFSoC. Embedding direct RF-sampling architecture into the MPSoC platform eliminates the need for discrete data converters and power-hungry serial interfaces (like JESD204) between the data converter and FPGA. In RFSoC, on-chip interface with the data converters is based on the Advanced eXtensible Interface (AXI) standard, AXI4-Stream, providing a high bandwidth and low-latency connection. Additionally, the programmable nature of this architecture provides hardware and software flexibility to a platform designed around RFSoC. 

Table~\ref{table:FPGA_TechProgression} shows the progression of FPGA technology over the last two decades, beginning with modest resources and capabilities of Virtex-4 and Virtex-5 FPGAs (used in the MWA digital receiver) to the state-of-the-art RFSoC device. For comparison, features and capabilities of XCKU9P Kintex UltraScale+ FPGA used in the TPM for SKA1-Low are also listed.

\begin{table*}[htb]
	\tabularfont
	\caption{Progression of FPGA technology}
	\begin{tabular}{lcccr}
		\topline
		FPGA Parameter&Virtex-4&Virtex-5 FPGA&Kintex UltraScale+&RFSoC Gen 3\\
		&(SX-35)&(SX-50T)&(XCKU9P)&(ZU49DR)\\
		\midline
		Introduced in&2004&2006&2015&2020\\
		Process technology&90-nm&65-nm&16-nm&16-nm\\
		Package-pins&FFG668&FFG665&FFVE900&F1760\\
		Package dimension(mm)&27x27&27x27&31x31&42.5x42.5\\
		Number of input/output pins&448&360&(96+208)&(96+312)\\
		System Logic cells&34.56K&52.22K&600K&930K\\
		Total Block RAM (Mb)&3.456&4.752&32.1&38\\
		Ultra RAM (Mb)&NA&NA&NA&22.5\\
		DSP slices&192(18x18)&288(25x18)&2520(27x18)&4272(27x18)\\
		PCI express blocks&--&--&--&2 PCIe Gen4 x8 lanes\\
		Number of transceivers, speed&--&12, up to 3.75~Gbps&28, 16.3~Gbps&16, up to 33~Gbps\\
		RF DAC&--&--&--&16x 14-bit 9.85~GSps DAC\\
		RF-sampling ADC&--&--&--&16x 14-bit 2.5~GSps  ADC\\
		\hline
	\end{tabular}
	\label{table:FPGA_TechProgression}
\end{table*}

On account of the skew between data and clock, it is an engineering challenge to interface the parallel digital output of a high-speed ADC with an FPGA. Further complications arise due to the assortment of ADC data output styles (parallel/serial/Single data rate/Double data rate) and standards. Due to the high bus speed of parallel ADCs, FPGAs use serializer/deserializer (SERDES) blocks to deserialize each bit in the bus to a wider and slower parallel data set. Deserializing allows the FPGA logic to handle wide buses at a comfortable clock speed. Afforded by the ever-shrinking geometries in process technology, ADC architecture has undergone a significant transformation with the (on-chip) integration of digital signal processing (DSP) features like: digital downconverter (DDC), numerically controlled oscillator (NCO), decimation filter, and JESD204 interface. These features have enhanced discrete ADC performance by bringing additional digital processing power. While the programmable nature of these features has provided flexibility in hardware and data interface, system design has benefited from a reduction in external components and design implementation time. To show the progression of ADC technology, Table~\ref{table:ADC_TechProgression} lists the specifications of RF-ADC inside the third-generation RFSoC and relevant specifications of modern direct RF-sampling ADCs used in digital receivers for MWA and LFAA. 

\begin{table*}[htb]
	\tabularfont
	\caption{Comparison of pertinent ADC parameters}
	\begin{tabular}{llcr}
		\topline
		ADC Parameter&AT84AD001C&AD9695BCPZ&ADCs in RFSoC Gen 3\\
		&(MWA digital-receiver)&(used in TPM)&(used in IPB)\\
		\midline
		Sampling clock (max.)&1~GSps&1.30~GSps&5~GSps\\
		Analog bandwidth&1.5~GHz&2~GHz&6~GHz\\
		ADC Resolution&8-bit&14-bit&14-bit\\
		Number of ADC channels&Dual ADC&Dual ADC&NA\\
		ADC digital output format&1:2 demux 8-bit parallel&Serial 4-lane JESD204B&AXI4-Stream parallel\\
		Power consumption/ADC&1.5~W&1.6~W&$\approx0.375$~W\\
		Package dimension(mm)&(22x22)&(9x9)&NA\\	
		Package-pins&LQPF144&64-Lead LFCSP&NA\\
		Process technology&400-nm BiCMOS&28-nm CMOS&16-nm (UltraScale+)\\
		\hline
	\end{tabular}
	\label{table:ADC_TechProgression}
\end{table*}

\begin{figure*}
	\centering\includegraphics[height= 9.75cm,width= 15.0cm]{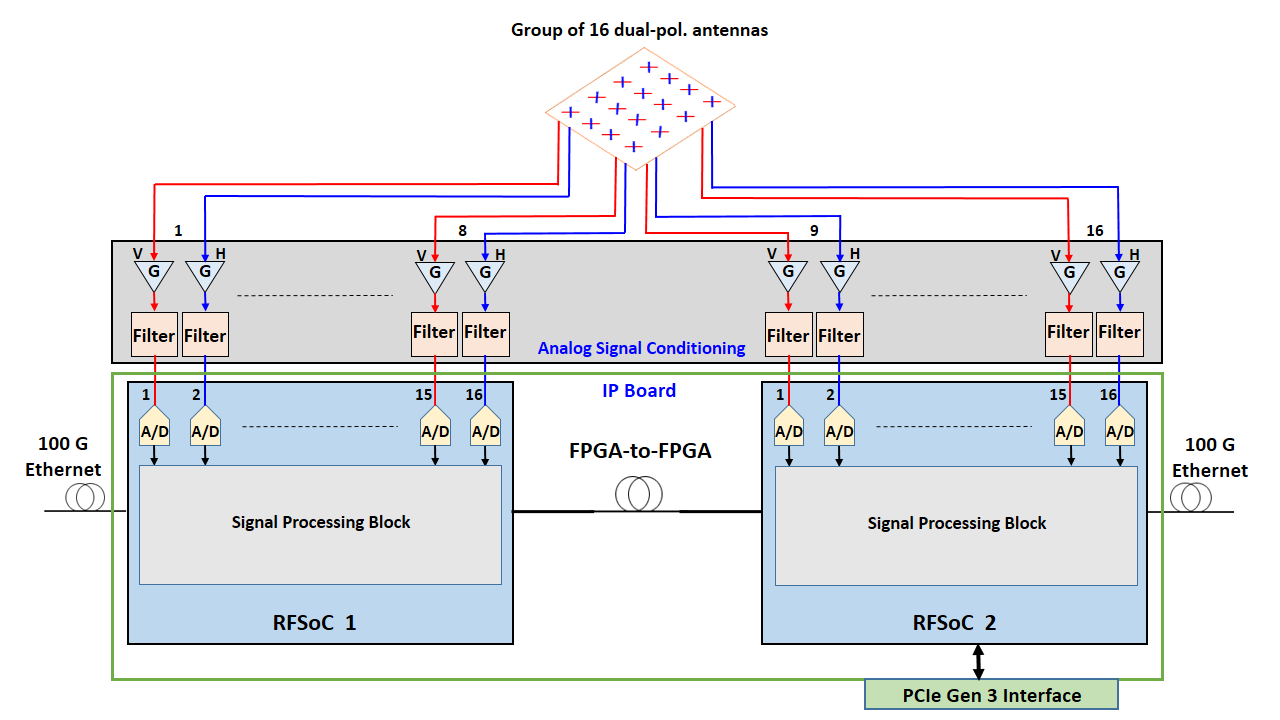}
	\caption{Within a compact footprint, the Integrated Prototype board implements a new generation digital-receiver architecture capable of 32 parallel analog-to-digital signal processing chains in two ZU49DR RFSoCs. IPB implements high-speed optical fiber interfaces for data transfer between RFSoCs and to the outside world.}
	\label{fig:IPB_BD}
\end{figure*}

At the Raman Research Institute, a high-speed signal processing platform, Integrated Prototype Board (IPB), is being designed and developed around two ZU49DR third-generation RFSoCs. IPB targets applications requiring multiple (up to 32) end-to-end analog-to-digital signal processing chains. Apart from the 16 on-chip high-performance 14-bit data converters, the ZU49DR RFSoC integrates a) 4200+ DSP slices for high-speed digital signal processing with a throughput of about 7500~GMACs$^{-1}$, b) multi-core processors, c) several high density (96) and high performance (312) user I/O pins, and d) sixteen 33~Gbps transceivers for high bandwidth serial I/O. IPB will be a Peripheral Component Interconnect Express (PCIe) full-length board with dimensions of 111.15~mm (height) x 312~mm (length). The Integrated Prototype Board will be replete with on-board DDR4 memory (up to 16 GB), optical 100~Gbps Ethernet connectivity between the two RFSoCs and RFSoC to the outside world, and 8x lanes PCIe Gen 3 connectivity with a data transfer rate of about 8~GBs$^{-1}$ for accelerated data processing\citep{aafreen2022high}. The feature-rich ZU49DR RFSoCs on IPB enables digitization of 32 RF paths and implementation of real-time signal processing like channelization, RFI detection and mitigation, and multi-beamforming within a compact footprint. Figure~\ref{fig:IPB_BD} shows a block diagram of the IPB.

In the design of the IPB, it was advantageous to use RFSoC instead of incorporating multiple discrete ADCs along with their interfaces to FPGA[s]. In addition to absorbing supporting ICs like power supply modules, clock generation and distribution chips, clock buffers, and decoupling capacitors and termination resistors, there is significant design complexity in implementing 16 discrete ADCs on a PCB. Table~\ref{table:ADC_TechProgression} lists the package dimension of the discrete ADC, AD9695BCPZ, used in the ADU (of TPM) to be 9~mm square. 16 ADCs and their associated circuits occupy a significant portion of the ADU's PCB real estate. Apart from the expensive real estate, routing high-speed PCB traces with minimal cross-coupling between them and maintaining power integrity and signal integrity in the entire PCB layer stack-up is challenging, time-consuming, and expensive. As RFSoC (42.5~mm square) subsumes most of these circuits, shrinking PCB dimension (up to 40~\%) reduces design time, cost, and implementation complexity. 16 dual-channel ADCs used in the ADU consume about 20~W. With the 16 higher performance RF-ADCs of ZU49DR estimated to consume about 6~W, there is $\approx70$\% reduction in ADC power consumption. In a situation requiring these high-performance ADCs to digitize analog paths of large-N aperture array (like the SKA-Low), a significant reduction in power requirements can be achieved using RFSoCs. JESD204B is a standardized high-speed serial data interface (intellectual property core) used by data converters to interface with logic devices (FPGAs or ASICs). JESD IP core may have to be purchased and instantiated inside an FPGA, incurring the cost, and consumption of valuable FPGA resources. As RFSoC implements the AXI4 protocol, design complexity and development time related to latency in the JESD protocol, complexity in routing high-speed length-matched PCB traces, and debugging the multi-lane interface for any faulty data transfer are avoided. Loading configuration sequence and M\&C of discrete ADCs require the implementation of a communication bus like I2C (Inter-Integrated Circuit) or SPI (Serial Peripheral Interface). As ADCs are integrated within an RFSoC, there is no need for an external communication bus, thus saving power and valuable real estate on the PCB. RFSoCs have an innovative lidless packaging design that allows for up to a 10°~C cooler operation for the same power dissipation as in a lidded package\footnote{\url{https://docs.xilinx.com/v/u/en-US/xapp1301-mechanical-thermal-design-guidelines}}.

In summary, ADCs subsumed into RFSoC provide power and area reduction, hardware and software flexibility, and scalability. There are also system design advantages concerning the generation and distribution of precision clock and synchronization signals. The IPB intends to leverage the above advantages to realize up to 32 end-to-end analog-to-digital signal processing chains as a power-efficient, compact form factor high-speed DSP platform for large-N radio astronomical applications.

\vspace{1em}
\section{Summary and Discussion}
\label{sec:SummaryDiscussion}

At the beginning of the new millennium, as technological advances enabled the construction of large aperture arrays capable of operating in the RFI-dominated low-frequency regime, there was a revival of low-frequency aperture array radio telescopes. Around that time, a plan to build the world’s largest radio telescope was gaining strength, culminating in several countries and organizations working together to design and construct the international SKA telescope. The Phase 1 project consisting of two telescopes, SKA1-Low and SKA1-Mid, will likely cost up to 2 billion euros for construction and the first ten years of operations. SKA pathfinder (including uGMRT) and precursor aperture array telescopes, operating in the frequency range of 10 to 300~MHz, have been built and operationalized to obtain valuable feedback on the instrument design, science derived, and operational aspects.

This review article primarily focused on how technological advancement has enabled the progression of digital-receiver architecture from MWA to SKA1-Low. Sections 2, 3, and 4 described the essential features of the digital receivers developed for the precursor Murchison Widefield Array, the upgraded GMRT (pathfinder), and the SKA1-Low telescope. Table~\ref{table:Summary_KeyDigRxParameters} presents a summary of key parameters of the digital receivers used in MWA, uGMRT, and SKA1-Low. Compared to the digital receivers for the MWA and uGMRT, Table~\ref{table:Summary_KeyDigRxParameters} clearly illustrates the scale, complexity, and challenges in implementing the digital receiver for SKA1-Low.  While the sampled bandwidth of the digital receiver for MWA and LFAA is comparable, the processed bandwidth of SKA1-Low (300~MHz) is an order of magnitude higher than MWA's 30~MHz. The processed bandwidth of uGMRT is 400~MHz maximum. Upgraded GMRT's and SKA1-Low's enhanced processed bandwidth is supported from the antennas, digital signal processing hardware, and data transport system to the final processing system that outputs science-quality data. The LFAA digital receiver, built around modern ADCs and high-performance FPGAs, complements the increased collecting area of SKA1-Low to make it a powerful instrument. Some of the LFAA receiver's capability enhancements and implications are discussed below.

\begin{table*}[htb]
	\tabularfont
	\caption{Summary of key parameters of the digital receiver used in MWA, uGMRT, and SKA1-Low}
	\begin{tabular}{llll} 
		\topline
		Digital receiver & MWA &  uGMRT & SKA1-Low\\
		parameter & digital receiver & digital receiver & digital receiver\\
		\midline
		Architecture&Built around 2 ADFBs &Built around iADCs &Built around 2 Pre-ADU\\
		&and 1 AgFo board & and ROACH FPGA boards & and 1 ADU boards = 1 TPM\\
		&&from CASPER&\\
		\\
		RF signals to be digitized&2 per tile x 128 tiles&2 per antenna x32 antennas&512 per station x512 stations\\
		(both polarization)& = 256&=64&=262144\\
		\\
		Number of Digital-receivers&16 Digital-receivers &32 iADC boards;& 16 TPMs per station;\\
		required&for 128 tiles&16 ROACH boards&8192 TPMs for 512 stations\\
		\\
		ADC details &8-bit, dual-channel&8-bit, dual channel & 14-bit, dual-channel\\
		&AT84AD001C (Atmel/e2v) & AT84AD001B (Atmel/e2v) & AD9695BCPZ (Analog Devices)\\
		\\
		Total number of&8 per Digital-receiver & 32 &16 ADCs per TPM;\\
		ADCs required & x16 receivers=128 & &131072 ADCs in 8192 TPMs\\
		\\
		Analog input signal & 80 – 300~MHz & Five different wave bands, & 50 – 350~MHz\\
		to ADC & &from 150~MHz to 1450~MHz &\\
		\\
		ADC sampling clock&655.36~MHz & 800~MHz & 800~MHz\\
		\\
		Sampled bandwidth;&327.68~MHz;  & 400~MHz;&400~MHz;\\
		Processed bandwidth & 30.72~MHz & 100, 200 or 400~MHz & 300~MHz\\
		\\
		Xilinx FPGA used& XCV4SX35 (Virtex-4), & XC5VSX95T (Virtex-5) & XCKU9P\\
		&XC5VSX50T (Virtex-5) & & (Kintex UltraScale+)\\
		\\
		Number of& 128 XCV4SX35 FPGAs;& 16 XC5VSX95T FPGAs & 2 XCKU9P FPGAs per TPM;\\
		FPGAs required & 16 XC5VSX50T FPGAs & &16384 FPGAs for 8192 TPMs\\
		\\
		Channelizer & 512-point critically- & Single-stage critically-sampled & 1024-point oversampled \\
		& sampled coarse PFB &  PFB, configurable from& (32/27) coarse PFB\\
		& & 4096- to 16384-points&\\
		\\
		Spectral resolution&1.28~MHz&12~kHz to 97~kHz,& 0.781~MHz\\
		&&for 200~MHz bandwidth&\\
		\\
		Type of beamformer&Analog beamformer&Digital beamformer in the & Digital beamformer in the\\
		& & frequency domain & frequency domain\\
		\\
		Number of beams formed & 1 beam & Up to 4 beams & Up to 48 beams\\
		\\
		Typical output data rate& 5.4 Gbps from & ~100 MB/s for incoherent addition & 23.2~Gbps from each ADU\\
		& each Digital-receiver & or phased array beamforming, & \\
		& & with 2048 spectral channels&\\
		\\
		Power consumption (typical) & $\approx75$ watts for 2 ADFB & $\approx90$ watts for 2 iADC boards & $\approx90$ watt for 1 ADU (16 dual-\\
		& boards and 1 AgFo board & and 1 ROACH FPGA board & channel ADCs and 2 FPGAs)\\
		\\
		\hline
	\end{tabular}
	\label{table:Summary_KeyDigRxParameters}
\end{table*}

\begin{itemize}
\item The MWA analog beamformer produces a tile beam that can be steered electronically. The analog beamformer has a fixed number of switchable delay units, so its resolution is limited. In the LFAA receiver chain, after accounting for length mismatches in the RFoF, there is a provision to adjust the geometric delay in multiples of the sampling clock period. As part of the frequency-domain beamforming process in the LFAA, residual phase correction is carried out on a per PFB output channel basis by multiplying with a suitable phase factor. This correction affords finer phase alignment between antennas across the array. The SKA1-Low implements digital beamforming in the Kintex UltraScale+ FPGA. The FPGA has sufficient resources allowing the formation of multiple beams (up to 48) on the sky and enhancing the total field of view of the telescope.

\item The 14-bit ADC used in the TPM provides better linearity performance than the 8-bit ADC used in the MWA digital receiver. It also provides a higher dynamic range performance and additional headroom for RFI.

\item The coarse PFB in the MWA digital receiver is not oversampled. Due to aliasing in the spectral channels of the coarse PFB, there is data loss as about 20-30\% of fine PFB channels have to be discarded. However, the signal processing chain in the SKA1-Low implements a two-stage PFB with an oversampled first stage and a critically-sampled second stage. Although this entails an increase in the speed of operation (by the oversampling factor of 1.18), the wider Nyquist bandwidth allows the fine PFB stage to extract the alias-free portion from each sub-band for further processing. A contiguous band, without data loss, can be obtained by stitching together alias-free spectral channels at the fine PFB output.
\end{itemize}

In summary, the SKA1-Low telescope, with its wider processed bandwidth, unsurpassed sensitivity due to a large collecting area, contiguous spectral coverage, flexibility in terms of the number of beams and bandwidth processed, and a host of receiver features, will be a powerful new instrument to address challenging problems in astronomy.

The upgraded GMRT’s back-end, a hybrid system combining FPGAs and CPU-GPUs, with similar capabilities as SKA1-Low, has been operating since 2017. It provides a pathway for building future large-N telescopes. The hybrid design allows shorter development times for implementing additional capabilities, e.g., RFI filtering, Raw-data recording, multiple narrowband modes, and offline processing.

The integration of RF-class data converters into MPSoC is a significant advancement in FPGA technology that can be leveraged for applications in radio astronomy. RFSoCs are ideally suited when there are a large number of antennas requiring end-to-end analog-to-digital signal processing chains. Using a system-on-chip that subsumes data converters provides advantages like reduction in power consumption, relaxation in system design challenges, and easing of implementation complexity. The availability of high-density phase-matched RF cable assemblies meant for multi-element designs and high-performance small-footprint optical cable systems for data transport at 10s of Gbps greatly simplifies IP board design. The IPB, owing to the above advantages, is well-suited for use in large-N applications.

The expanded GMRT (eGMRT) is a proposal to equip GMRT with focal plane arrays (FPAs) for enhanced field of view \citep{patra2019expanded}. A compact RFSoC-based digital receiver providing a) direct RF digitization with 14-bit ADCs, b) hardware features for high-speed real-time digital signal processing, and c) several GTY transceivers for large bandwidth I/O can be considered for eGMRT. Coupling of self-generated RFI from the digital receiver to the signal path can be significantly reduced by mounting it inside an RF shielding enclosure\citep{girish2020saras}. After digitization and first-stage digital signal processing, aggregated data can be packetized and transported to downstream stages for further processing, via high-speed optical fiber links implemented on the RFSoC platform. Similarly, as a SKA-Low station incorporates a large number of antennas, an RFSoC-based platform could be advantageous for Phase 2 of SKA. Such a platform in a compact form factor eliminates several discrete components, minimizes clock distribution and synchronization challenges, reduces power consumption, and addresses latency issues through the on-chip interface between ADC output and the first-stage DSP blocks.

\section*{Acknowledgements}
We thank Raman Research Institute (RRI) for supporting the Integrated Prototype Board activities taking place under the SKA India Consortium (SKAIC) project proposal (DPR, 2019). Authors Girish, Srivani, and Prabu thank Manjunath Bevinamar, Hardik Purohit, and Jeetesh Tiwari at Avnet India Pvt. Ltd. for providing valuable insights into Gen 3 RFSoCs and enthusiastically supporting us during IPB activities. Thanks to Shirisha V. at RRI for proofreading.
\vspace{-1em}

\setlength{\tabcolsep}{20pt}
\setlength{\columnsep}{2cm}
\renewcommand{\arraystretch}{1.5}


\vspace{1.5em}
\newcommand{\comment}[1]{}
\comment{

}
\bibliography{bibliography}

\begin{thebibliography}{}
\expandafter\ifx\csname natexlab\endcsname\relax\def\natexlab#1{#1}\fi

\bibitem[{Aafreen {$et~al$.}(2022)Aafreen, Abhishek, Ajithkumar, Vaidyanathan,
  Barve, Bhattramakki, Bhat, Girish, Ghalame, Gupta,
  {$et~al$.}}]{aafreen2022high}
Aafreen, R., Abhishek, R., Ajithkumar, B., {$et~al$.} 2022, arXiv preprint
  arXiv:2207.07054, 21

\bibitem[{Bellanger \& Daguet(1974)}]{bellanger1974tdm}
Bellanger, M., \& Daguet, J. 1974, IEEE Transactions on Communications, 22, 7

\bibitem[{Buch {$et~al$.}(2016)Buch, Bhatporia, Gupta, Nalawade, Chowdhury,
  Naik, Aggarwal, \& Ajithkumar}]{buch2016towards}
Buch, K.~D., Bhatporia, S., Gupta, Y., {$et~al$.} 2016, Journal of Astronomical
  Instrumentation, 5, 14

\bibitem[{Caputa {$et~al$.}(2022)Caputa, Harrison, Ljusic, Pleasance, \&
  Zhao}]{caputa2022ska}
Caputa, K., Harrison, S., Ljusic, Z., Pleasance, M., \& Zhao, E. 2022, in
  Ground-based and Airborne Telescopes IX, Vol. 12182, Proc. of SPIE, p. 19

\bibitem[{Comoretto {$et~al$.}(2017)Comoretto, Chiello, Roberts, Halsall,
  Adami, Alderighi, Aminaei, Baker, Belli, Chiarucci,
  {$et~al$.}}]{comoretto2017signal}
Comoretto, G., Chiello, R., Roberts, M., {$et~al$.} 2017, Journal of
  Astronomical Instrumentation, 6, 17

\bibitem[{Comoretto {$et~al$.}(2020)Comoretto, Monari, Belli, Chiarucci,
  Schillir{\`o}, Schiaffino, Perini, Mattana, Alderighi, d'Angelo,
  {$et~al$.}}]{comoretto2020signal}
Comoretto, G., Monari, J., Belli, C., {$et~al$.} 2020, in Ground-based and
  Airborne Telescopes VIII, Vol. 11445, Proc. of SPIE, p. 14

\bibitem[{De \& Gupta(2016)}]{de2016real}
De, K., \& Gupta, Y. 2016, Experimental Astronomy, 41, 25

\bibitem[{de~Lera~Acedo {$et~al$.}(2015)de~Lera~Acedo, Razavi-Ghods, Troop,
  Drought, \& Faulkner}]{de2015skala}
de~Lera~Acedo, E., Razavi-Ghods, N., Troop, N., Drought, N., \& Faulkner, A.
  2015, Experimental Astronomy, 39, 28

\bibitem[{DeBoer {$et~al$.}(2017)DeBoer, Parsons, Aguirre, Alexander, Ali,
  Beardsley, Bernardi, Bowman, Bradley, Carilli,
  {$et~al$.}}]{deboer2017hydrogen}
DeBoer, D.~R., Parsons, A.~R., Aguirre, J.~E., {$et~al$.} 2017, Publications of
  the Astronomical Society of the Pacific, 129, 27

\bibitem[{Deller {$et~al$.}(2007)Deller, Tingay, Bailes, \&
  West}]{deller2007difx}
Deller, A.~T., Tingay, S., Bailes, M., \& West, C. 2007, Publications of the
  Astronomical Society of the Pacific, 119, 19

\bibitem[{Dewdney {$et~al$.}(2009)Dewdney, Hall, Schilizzi, \&
  Lazio}]{dewdney2009square}
Dewdney, P.~E., Hall, P.~J., Schilizzi, R.~T., \& Lazio, T. J.~L. 2009,
  Proceedings of the IEEE, 97, 15

\bibitem[{Garrett(2013)}]{garrett2013radio}
Garrett, M.~A. 2013, in 2013 Africon Conference, IEEE Xplore, p. 5

\bibitem[{Girish {$et~al$.}(2020)Girish, Srivani, Subrahmanyan, Udaya~Shankar,
  Singh, Jishnu~Nambissan, Sathyanarayana~Rao, Somashekar, \&
  Raghunathan}]{girish2020saras}
Girish, B.~S., Srivani, K.~S., Subrahmanyan, R., {$et~al$.} 2020, Journal of
  Astronomical Instrumentation, 9, 17

\bibitem[{Gupta(2014)}]{gupta2014gmrt}
Gupta, Y. 2014, The Metrewavelength Sky, 13, 7

\bibitem[{Gupta {$et~al$.}(2017)Gupta, Ajithkumar, Kale, Nayak, Sabhapathy,
  Sureshkumar, Swami, Chengalur, Ghosh, Ishwara-Chandra, {$et~al$.}}]{gupta}
Gupta, Y., Ajithkumar, B., Kale, H., {$et~al$.} 2017, Current Science, 113, 8

\bibitem[{Harris {$et~al$.}(2003)Harris, Dick, \& Rice}]{harris2003digital}
Harris, F.~J., Dick, C., \& Rice, M. 2003, IEEE transactions on microwave
  theory and techniques, 51, 18

\bibitem[{Lonsdale {$et~al$.}(2009)Lonsdale, Cappallo, Morales, Briggs,
  Benkevitch, Bowman, Bunton, Burns, Corey, DeSouza,
  {$et~al$.}}]{lonsdale2009murchison}
Lonsdale, C.~J., Cappallo, R.~J., Morales, M.~F., {$et~al$.} 2009, Proceedings
  of the IEEE, 97, 9

\bibitem[{Manley \& Kapp(2012)}]{manley2012meerkat}
Manley, J., \& Kapp, F. 2012, in 2012 International Conference on
  Electromagnetics in Advanced Applications, IEEE, p. 4

\bibitem[{Morrison {$et~al$.}(2020)Morrison, Bunton, van Straten, Deller, \&
  Jameson}]{morrison2020performance}
Morrison, I., Bunton, J., van Straten, W., Deller, A., \& Jameson, A. 2020,
  Journal of Astronomical Instrumentation, 9, 17

\bibitem[{Naldi {$et~al$.}(2017)Naldi, Mattana, Pastore, Alderighi, Zarb~Adami,
  Schillir{\`o}, Aminaei, Baker, Belli, Comoretto,
  {$et~al$.}}]{naldi2017digital}
Naldi, G., Mattana, A., Pastore, S., {$et~al$.} 2017, Journal of Astronomical
  Instrumentation, 6, 17

\bibitem[{Parsons {$et~al$.}(2010)Parsons, Backer, Foster, Wright, Bradley,
  Gugliucci, Parashare, Benoit, Aguirre, Jacobs,
  {$et~al$.}}]{parsons2010precision}
Parsons, A.~R., Backer, D.~C., Foster, G.~S., {$et~al$.} 2010, The Astronomical
  Journal, 139, 13

\bibitem[{Patra {$et~al$.}(2019)Patra, Kanekar, Chengalur, Sharma, de~Villiers,
  Ajit~Kumar, Bhattacharyya, Bhalerao, Bombale, Buch,
  {$et~al$.}}]{patra2019expanded}
Patra, N.~N., Kanekar, N., Chengalur, J.~N., {$et~al$.} 2019, Monthly Notices
  of the Royal Astronomical Society, 483, 16

\bibitem[{Perley {$et~al$.}(2009)Perley, Napier, Jackson, Butler, Carlson,
  Fort, Dewdney, Clark, Hayward, Durand, {$et~al$.}}]{perley2009expanded}
Perley, R., Napier, P., Jackson, J., {$et~al$.} 2009, Proceedings of the IEEE,
  97, 15

\bibitem[{Prabu {$et~al$.}(2014)Prabu, Srivani, Kamini, Madhavi, Deshpande,
  Udaya~Shankar, Subrahmanyan, Briggs, Roshi, Ord, {$et~al$.}}]{prabu2014full}
Prabu, T., Srivani, K., Kamini, P., {$et~al$.} 2014, in Astronomical Society of
  India Conference Series, Vol.~13, p. 5

\bibitem[{Prabu {$et~al$.}(2015)Prabu, Srivani, Roshi, Kamini, Madhavi, Emrich,
  Crosse, Williams, Waterson, Deshpande, {$et~al$.}}]{prabu2015digital}
Prabu, T., Srivani, K., Roshi, D.~A., {$et~al$.} 2015, Experimental Astronomy,
  39, 21

\bibitem[{Reddy {$et~al$.}(2017)Reddy, Kudale, Gokhale, Halagalli, Raskar, De,
  Gnanaraj, Ajith~Kumar, \& Gupta}]{reddy2017wideband}
Reddy, S.~H., Kudale, S., Gokhale, U., {$et~al$.} 2017, Journal of Astronomical
  Instrumentation, 6, 16

\bibitem[{Roy {$et~al$.}(2010)Roy, Gupta, Pen, Peterson, Kudale, \&
  Kodilkar}]{roy2010real}
Roy, J., Gupta, Y., Pen, U.-L., {$et~al$.} 2010, Experimental Astronomy, 28, 36

\bibitem[{Schilizzi {$et~al$.}(2008)Schilizzi, Dewdney, \&
  Lazio}]{schilizzi2008square}
Schilizzi, R.~T., Dewdney, P.~E., \& Lazio, T. J.~W. 2008, in Ground-based and
  Airborne Telescopes II, Vol. 7012, Proc. of SPIE, p. 13

\bibitem[{Schillir{\`o} {$et~al$.}(2020)Schillir{\`o}, Alderighi, Belli,
  Chiarucci, Chiello, Comoretto, D'Angelo, Magro, Mattana, Monari,
  {$et~al$.}}]{schilliro2020design}
Schillir{\`o}, F., Alderighi, M., Belli, C., {$et~al$.} 2020, in Ground-based
  and Airborne Telescopes VIII, Vol. 11445, Proc. of SPIE, p. 16

\bibitem[{Schinckel {$et~al$.}(2012)Schinckel, Bunton, Cornwell, Feain, \&
  Hay}]{schinckel2012australian}
Schinckel, A.~E., Bunton, J.~D., Cornwell, T.~J., Feain, I., \& Hay, S.~G.
  2012, in Ground-based and Airborne Telescopes IV, Vol. 8444, Proc. of SPIE,
  p. 12

\bibitem[{Tingay {$et~al$.}(2013)Tingay, Goeke, Bowman, Emrich, Ord, Mitchell,
  Morales, Booler, Crosse, Wayth, {$et~al$.}}]{tingay2013murchison}
Tingay, S.~J., Goeke, R., Bowman, J.~D., {$et~al$.} 2013, Publications of the
  Astronomical Society of Australia, 30, 21

\bibitem[{van Haarlem {$et~al$.}(2013)van Haarlem, Wise, Gunst, Heald, McKean,
  Hessels, de~Bruyn, Nijboer, Swinbank, Fallows, {$et~al$.}}]{van2013lofar}
van Haarlem, M.~P., Wise, M.~W., Gunst, A., {$et~al$.} 2013, Astronomy \&
  astrophysics, 556, 53

\bibitem[{Wayth {$et~al$.}(2018)Wayth, Tingay, Trott, Emrich, Johnston-Hollitt,
  McKinley, Gaensler, Beardsley, Booler, Crosse, {$et~al$.}}]{wayth2018phase}
Wayth, R.~B., Tingay, S.~J., Trott, C.~M., {$et~al$.} 2018, Publications of the
  Astronomical Society of Australia, 35, 9

\end{thebibliography}

\end{document}